\documentclass[aps,prx,reprint,
superscriptaddress,floatfix,longbibliography,footinbib]{revtex4-1}
\usepackage{mathtools}
\usepackage{amsfonts}
\usepackage{amsmath}
\usepackage{amssymb}
\usepackage{physics}
\usepackage{float}
\usepackage{bbold}
\usepackage[version=4]{mhchem}
\usepackage{xcolor}
\usepackage{soul}
\usepackage{siunitx}
\usepackage{tabularray}

\newcommand{\effcoupling}{\lambda} 
\newcommand{\Deltaexdressed}{\Delta_{\text{ex}}^D} 
\newcommand{\Deltacdressed}{\Delta_c^D}
\newcommand{\kappadressed}{\kappa^D} 
\AtBeginDocument{\RenewCommandCopy\qty\SI}
\DeclareSIUnit\angstrom{\text{\AA}}
\DeclareSIUnit\elementarycharge{e}

\begin{document}

\title{Raman-phonon-polariton condensation in a transversely pumped cavity}

\author{Alexander N.~Bourzutschky}
\affiliation{Department of Physics, Stanford University, Stanford CA 94305, USA}
\affiliation{E.~L.~Ginzton Laboratory, Stanford University, Stanford, CA 94305, USA}
\author{Benjamin L.~Lev}
\affiliation{Department of Physics, Stanford University, Stanford CA 94305, USA}
\affiliation{E.~L.~Ginzton Laboratory, Stanford University, Stanford, CA 94305, USA}
\affiliation{Department of Applied Physics, Stanford University, Stanford CA 94305, USA}
\author{Jonathan Keeling}
\affiliation{SUPA, School of Physics and Astronomy, University of St.~Andrews, St. Andrews KY16 9SS, United Kingdom}

\date{\today}

\begin{abstract}

Phonon polaritons are hybrid states of light and matter that are typically realised when optically active phonons couple strongly to photons.  We suggest a new approach to realising phonon polaritons, by employing a transverse-pumping Raman scheme, as used in experiments on cold atoms in optical cavities. This approach allows hybridisation between an optical cavity mode and any Raman-active phonon mode. Moreover, this approach enables one to tune the effective phonon--photon coupling by changing the strength of the transverse pumping light. We show that such a system may realise a phonon-polariton condensate. To do this, we find the stationary states and use Floquet theory to determine their stability. We thus identify distinct superradiant and lasing states in which the polariton modes are macroscopically populated. We map out the phase diagram of these states as a function of pump frequencies and strengths. Using parameters for transition metal dichalcogenides, we show that realisation of these phases may be practicably obtainable. The ability to manipulate phonon mode frequencies and attain steady-state populations of selected phonon modes provides a new tool for engineering correlated states of electrons.

\end{abstract}

\maketitle

\section{Introduction}

Phonon polaritons result from strong coupling between light and lattice vibrations, leading to new normal modes with hybrid properties of both constituents~\cite{Basov2020pp}. Because phonons play an essential role in material properties---including mediating phases such as superconductivity---manipulating phonon properties by coupling to light is an intriguing prospect~\cite{Sentef2018cqp}. However, phonons are typically in the terahertz (THz) or infrared range, and direct strong coupling requires cavities supporting THz or infrared cavity modes.  Moreover, only optically active phonon modes---i.e., those with a dipole matrix element---can couple directly to light. In many cases, relevant phonon modes may not be optically active; see e.g.~\cite{Mankowsky2014nld}. As we demonstrate further below, this restriction can be overcome by using a Raman pumping scheme.  That is, one couples to the phonons via a two-photon transition involving cross terms between cavity photons and an external pump. This engineers an effective coupling between optical cavity photons and terahertz phonons. 

Raman-driving schemes~\cite{Dimer2007pro} have proven highly fruitful for experiments involving cold atoms in optical cavities.  Crucially, such schemes realise a scenario that may be referred to as ``synthetic cavity QED,'' in which one can control the strength of matter-light coupling as well as the effective energies of the matter and light degrees of freedom.  Following the original proposal~\cite{Dimer2007pro}, this enabled the realisation~\cite{Baumann2010dqp} of the Dicke-Hepp-Lieb superradiance transition~\cite{Hepp1973ots,Wang1973pti,Kirton2019itt}.  Further extensions of this core idea of Raman driving to multimode confocal cavities~\cite{Kollar2015aac} have introduced tunable-range~\cite{Vaidya2018tpa} and sign-changing interactions~\cite{Guo2019spa} to cavity QED, enabling the realisation of an optical lattice with sound modes~\cite{Guo2021aol} and replica-symmetry breaking in a spin glass~\cite{Kroeze2023rsb}.
See the reviews~\cite{Ritsch2013cai,Mivehvar2021cqw} for further discussion. Our aim in this manuscript is to explore how the power and flexibility of Raman-driving can be extended to solid-state systems.

As noted above, being able to manipulate the properties of solid-state phonon modes is of significant interest due to the possibility of engineering correlated electronic phases. Using transient light pulses to modify phonon modes has been central to proposals for, and realisations of, light-induced phase transitions in cuprates and other superconductors ~\cite{Fausti2011lsi,Mankowsky2014nld,Mitrano2016pls,Cavalleri2017ps}.  Such results rely on direct, terahertz coupling of light to the phonons in the transient, pulsed-laser regime, and thus inevitably causes heating while posing challenges to those attempting to interpret the results.  This has led to proposals to instead use strong coupling to infrared or THz cavities~\cite{Curtis2022cmi,schlawin2022cqm}, including a variety of proposals of how cavity QED may enhance superconductivity~\cite{Schlawin2019ces,Curtis2019cqe,Gao2020pep}. Even with a cavity, coupling is restricted to optically active modes. To overcome this, Raman driving schemes are required; see Ref.~\cite{collado2023epa} for an alternate Raman scheme providing parametric coupling between a THz photon and phonon.

In this paper, we show that it is possible to achieve a superradiant phase---i.e.,~a phonon-polariton condensate---using a continuous-wave Raman driving scheme coupling a phonon to an optical-frequency cavity mode. We present a model for independent, driven excitons in an optical cavity and use the mean-field ansatz to show that sufficiently strong driving leads to an instability of the normal phase, understood to be the phase with no macroscopic population of the cavity. We show that the instability can take two forms, either due to superradiance or due to lasing, and we describe the features of each. We present the phase diagram for the model and discuss the effect of different parameters.

Our model is inspired by transition metal dichalcogenides (TMDs), particularly TMD monolayers; these have attracted much attention for their optoelectronic properties~\cite{Koppens2014pbo}. Key to the results below is that they host sharp~\cite{Cadiz2017ela}, strongly bound, optically active excitons (electron-hole pairs)~\cite{Li2014mot,Wang2018iei} that couple to sufficiently long-lived optical phonons with a Huang-Rhys parameter of approximately unity~\cite{Li2021ecs,Kumar2022ect}. The model presented in this paper is however not limited to TMD materials, but presents a general approach for how Raman-driving may be used in solid-state contexts. Hence the results below may be applied to more exotic few-layer and twisted multilayer materials.


\section{Results}
\subsection{Model: Phonons and Driven Excitons in a Cavity}
\label{sec:model}

\begin{figure}
    \centering
    \includegraphics[width=\linewidth]{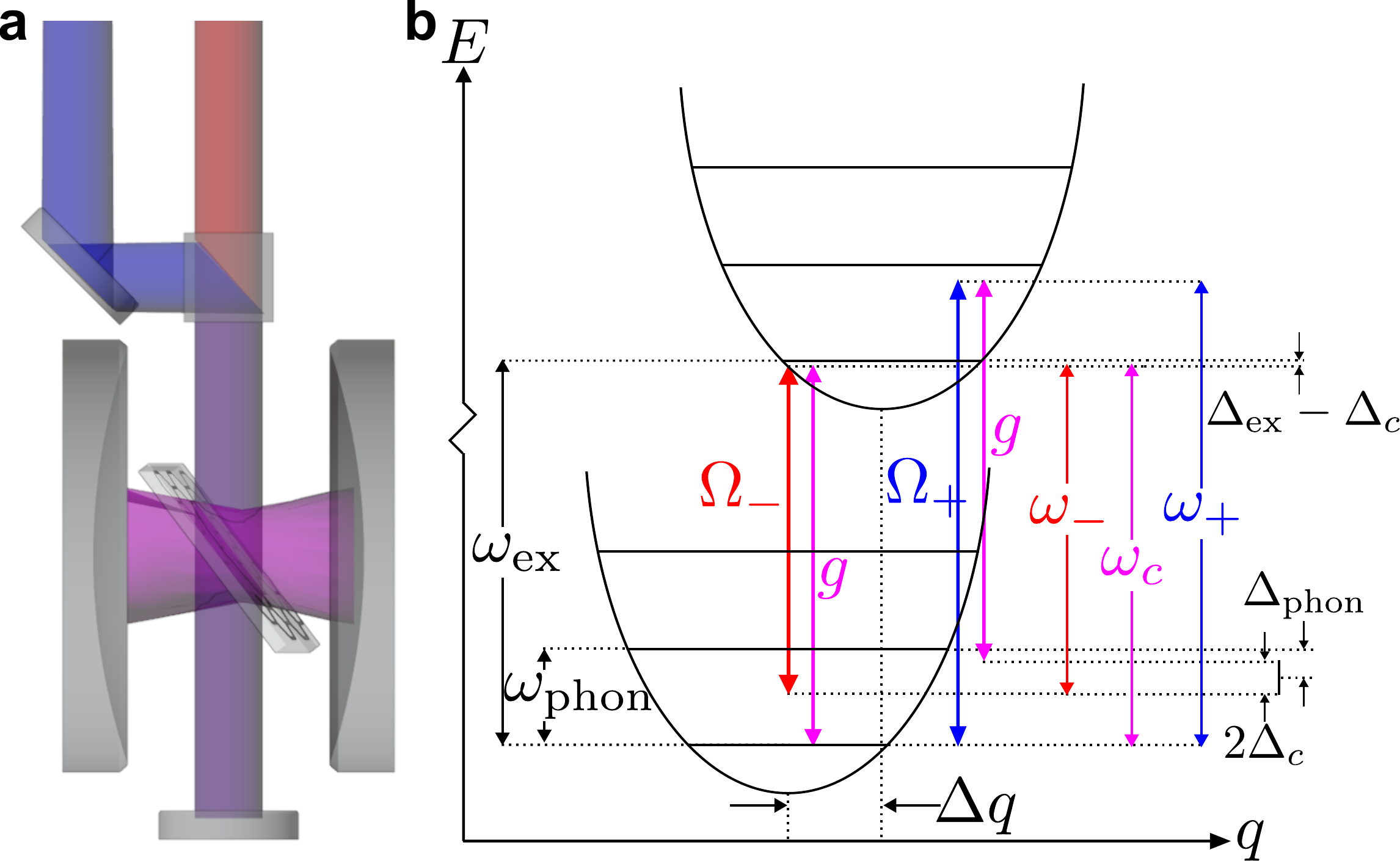}
    \caption{(a) Schematic of the scenario considered. Two pumps (red and blue beams) drive a two-dimensional material coupled to an optical cavity mode (pink). (b) Energy-level diagram showing the two electronic states, vibrational sub-bands and associated pump strengths and energies.  For definitions of detunings, see Table~\ref{tab:detunings}.}
    \label{fig:schematic_levels}
\end{figure}

We consider a model that involves the interaction of cavity photons $\hat a$, excitons $\hat X_j$, and phonons $\hat b_j$. Our model considers explicit exciton-photon coupling (including optical driving of the excitons by the pumps) and exciton-phonon coupling.  As discussed further below, this leads to an effective Raman coupling between phonons and photons; however, in contrast to other Raman-driven schemes~\cite{Dimer2007pro}, we will not adiabatically eliminate the excitons.

To simplify the calculation, we consider the case of a single-mode optical cavity. We will remark later on how these results might change in a multimode cavity. Furthermore, we consider a model of quasilocalised excitons on sites labelled $j$, each with an associated phonon mode.
Setting $\hbar=1$ throughout, this model takes the form
$\hat{H}=\omega_c \hat{a}^\dagger \hat{a} + \sum_{j=1}^N \hat{H}_{\text{mat},j}$, where the matter Hamiltonian for each site $j$ is
\begin{multline}
    \label{eq:Horig}
    \hat{H}_{\text{mat},j}=
    \omega_{\text{ex}} \hat{X}^\dagger_j \hat{X}_j + \omega_{\text{phon}} 
    \left[\hat{b}^\dagger_j \hat{b}_j  + \Delta q \left( \hat{b}^\dagger_j + \hat{b}_j \right) \hat{X}^\dagger_j \hat{X}_j \right] \\
    + \left( \hat{X}^\dagger_j + \hat{X}_j \right)  \left[ \Omega_+ \frac{e^{i\omega_+ t}}{2} + \Omega_- \frac{e^{i \omega_- t}}{2} + g_0 \hat{a}^\dagger
    + \text{H.c.} \right].
\end{multline}
We note that $\Delta q$ parameterises the exciton-phonon coupling, which takes the Holstein form~\cite{Holstein1959sop}.  The Huang-Rhys parameter $S$~\cite{Huang1950tol}, which is typically used to define coupling to photons, is related to $\Delta q$ by $S=\Delta q^2$. The terms $\Omega_{\pm} e^{i \omega_\pm t}$ denote the two driving fields.  As discussed elsewhere~\cite{Dimer2007pro,Kroeze2018sso}, two-frequency driving is used to ensure resonance of both co- and counter-rotating couplings between photons and phonons, as illustrated in Fig.~\ref{fig:schematic_levels}. That is, there are two processes we want to make resonant: The process that destroys a photon and creates a phonon (the co-rotating process), and the process that creates both a photon and a phonon (the counter-rotating process).  When using a single pump frequency, the pump can be chosen to make one of these two-photon transitions resonant, but not both.  Using two pumps allows for the simultaneous two-photon resonance of both processes.  As we will discuss further below, this simultaneous resonance can reduce the threshold power required by absorbing the phonon energy into the pumps' difference frequency.

In addition to the Hamiltonian, we consider effects of dissipation by considering a Lindblad master equation~\cite{Breuer2007tto}:
$i \partial_t \rho = [\hat{H},\rho] + i \sum_j \mathcal{D}[\hat{L}_j]$, with $\mathcal{D}[\hat{L}]=
\hat{L} \rho  \hat{L}^\dagger -  \{ \hat{L}^\dagger \hat{L},\rho\}/2$. We consider dissipation corresponding to cavity loss $\hat{L}_c=\sqrt{2 \kappa} \hat{a}$, exciton loss $\hat{L}_{\text{ex},j}=\sqrt{\Gamma_{\text{ex}}} \hat{X}_j$, and phonon loss $\hat{L}_{b,j}=\sqrt{\Gamma_{\text{phon}}} \hat{b}_j$.
The symbol definitions and values used are included in Table~\ref{tab:parameters}. 

The cavity and pump frequencies are chosen to be close to that of the exciton, and so we move into a frame rotating at the mean pump frequency $\overline{\omega}$ and disregard counter-rotating terms. We then implement a mean-field ansatz (details given in Methods) and obtain coupled equations for the rescaled cavity field $\alpha=\expval{\hat{a}}/\sqrt{N}$ and the density matrix of a single site $\rho$:
\begin{align}
    \label{eq:mf1}
    \partial_t \alpha &= -i \left[ -\Delta_c \alpha + g_0 \sqrt{N}\, \text{Tr} \left( \rho \hat{X} \right) \right] - \kappa \alpha; \\
    \label{eq:mf2}
    \partial_t \rho &= -i \left[ \hat{H}^{\text{MF}}_{\text{mat}},  \rho \right] + \mathcal{D}\left[ \sqrt{\Gamma_{\text{ex}}} \hat{X} \right] + \mathcal{D}\left[ \sqrt{\Gamma_{\text{phon}}} \hat{b} \right].
\end{align}
The mean-field material Hamiltonian $\hat{H}^{\text{MF}}_{\text{mat}}$ is
\begin{multline}
    \label{eq:mfHmat}
    \hat{H}^{\text{MF}}_{\text{mat}} = -\Delta_{\text{ex}} \hat{X}^\dagger \hat{X} + \omega_{\text{phon}} \left[\hat{b}^\dagger \hat{b} +  \Delta q \left(\hat{b} + \hat{b}^\dagger\right) \hat{X}^\dagger \hat{X} \right]\\
    + \left[ \left( \Omega_{+} \frac{ e^{-i \Delta\omega t}}{2} + \Omega_{-} \frac{ e^{+i \Delta\omega t}}{2} + g_0 \sqrt{N} \alpha \right) \hat{X}^\dagger + \text{H.c.} \right],
\end{multline}
where the detunings are given in Table~\ref{tab:detunings}.

\begin{table}[b] 
\centering
\caption{Frequencies and detunings used in the transverse pumping model.}
\begin{tblr}{width=0.48\textwidth,colspec={Q[r,m]m{1em}Q[l,m]X[c,m]}} \hline \hline
Symbol && Definition  & Meaning \\ \hline
$\overline{\omega}$ &$\equiv$&$\frac{1}{2} \left( \omega_+ + \omega_- \right)$ & Mean pump frequency  \\
$\Delta_{\text{ex}}$ &$\equiv$& $\overline{\omega} - \omega_{\text{ex}}$& Exciton detuning  \\
$\Delta_c$ &$\equiv$& $\overline{\omega} - \omega_c$ & Cavity detuning  \\
$\Delta \omega$ &$\equiv$& $\frac{1}{2} \left( \omega_+ - \omega_- \right)$& Half of the splitting of pump frequencies  \\
$\Delta_{\text{phon}}$ &$\equiv$&$\Delta\omega - \omega_{\text{phon}}$ & Detuning from $\Delta\omega$ to the phonon energy  \\ \hline
$\Deltaexdressed$ & $\equiv$ & $\Delta_{\text{ex}} + (\Delta q)^2 \omega_{\text{phon}} $ & Dressed exciton detuning  \\
$\Deltacdressed$ & $\equiv$ & $\Delta_c - \frac{g_0^2 N}{\Deltaexdressed - \Delta_c}$ & Dressed cavity detuning  \\
\hline \hline
\end{tblr} \label{tab:detunings}
\end{table}

There remains explicit time dependence in the exciton drive at frequency $\Delta\omega$.  As noted above, and can be seen in Fig.~\ref{fig:schematic_levels}(b), this detuning is included so that there can be resonant processes for scattering a phonon into a photon, as well as resonant processes to create a phonon-photon pair.  Such resonance occurs if $\Delta_{\text{phon}}\equiv\Delta\omega - \omega_{\text{phon}}=0$. In most prior work on Raman-driving schemes, the exciton state would be eliminated, and the above resonant processes would then lead to effective couplings $(\hat{a}+\hat{a}^\dagger)(\hat{b}_j+\hat{b}_j^\dagger)$.  This adiabatic elimination is not so clearly justified in this work, as we cannot necessarily operate in a parameter regime where the exciton detuning is larger than all other detunings---we discuss relevant parameter values in the next subsection.

\begin{figure}[t!]
    \centering
    \includegraphics[width=\linewidth]{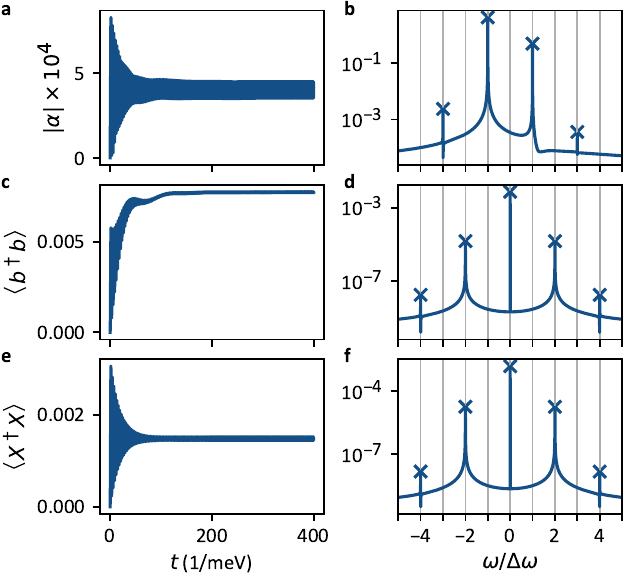}
    \caption{Comparison between direct time evolution of Eqs.~\eqref{eq:mf1} and~\eqref{eq:mf2} and the stationary-state solution in the form of a Fourier series.  Shown are three different observables:  (a,b) Photon field; (c,d) phonon number; and (e,f) exciton number.  Left plots show the time evolution. Right plots show, on a logarithmic scale, the magnitudes of the discrete Fourier transforms of the late-time evolution shown as lines from $900$ to $\qty[per-mode=symbol]{1000}{\per\meV}$ compared to the Fourier components present in the stationary-state solution shown as crosses.  For the time-evolution results, each peak is broadened due to the finite duration of the simulation. Parameter values are $\Delta_c = \qty{-0.25}{\meV}$ and $\Omega_\pm = \qty{0.2}{\meV}$, along with those listed in Table~\ref{tab:parameters}; the system is initialised in the vacuum state.}
    \label{fig:compare-te-ss}
\end{figure}

In the following, we will solve this periodically driven and dissipative mean-field model.  Due to the explicit time dependence, there are no true steady-state solutions.  However, in many cases the solution is commensurate with the drive frequency.  We will refer to such solutions as \textit{stationary states}. To avoid requiring us to perform direct time evolution of Eqs.~\eqref{eq:mf1} and \eqref{eq:mf2}, we can directly find the stationary states by looking for solutions in the form of a Fourier series: $\alpha(t) = \sum_n \alpha_{\text{ss}, n} e^{i n \Delta \omega t},$ and $\rho(t) = \sum_n \rho_{\text{ss}, n} e^{i n \Delta \omega t}$, where the sum over $n$ is truncated at some maximum $|n|$. Figure~\ref{fig:compare-te-ss} shows that this procedure well matches the results of direct time evolution. See Methods for details.

The stationary-state solver always finds a normal state with $\alpha_{\text{ss},0}=0$, i.e., one with zero steady population of the cavity.  However, because the pumps are still interacting with excitons and excitons continue to scatter into the cavity, the nonzero-frequency components of $\alpha_{\text{ss}}$ in the normal state do not vanish---specifically terms with odd $n$ are nonzero, as seen in Fig.~\ref{fig:compare-te-ss}. While this normal state always exists, it is not always stable.  To determine this, we linearise the master equation about one of its stationary states.  This results in an equation in Floquet form. The Floquet exponents found from this equation then determine the stability of the stationary state: If any Floquet exponent has real part greater than zero, then the state is unstable. For our time-dependent system, these Floquet exponents play the role that linear-stability eigenvalues played in previous work~\cite{Bhaseen2012don}. We therefore refer to these Floquet exponents as eigenvalues in what follows; see Methods for further details. Regimes where the normal state with $\alpha_{\text{ss},0}=0$ is stable we term the \textit{normal phase}. Regimes where the normal state is unstable are discussed below, as we find multiple other states that may arise.

\subsection{Low-Energy Effective Model}
\label{sec:leem}

Before presenting results found directly from the stationary state and stability analysis of the full photon-exciton-phonon model discussed in the previous section, we introduce an effective model that provides a qualitative understanding of the results.  This effective model results from adiabatic elimination of the exciton states, leading to an effective photon-phonon model, as discussed in the Supplementary Information~\cite{supp}.   It takes the form
$\hat{H}_{\text{eff}} = -\Deltacdressed \hat{a}^\dagger \hat{a} + \sum_{j = 1}^N \hat{H}_{\text{eff}, j}$, with
\begin{equation}
    \label{eq:H_effj}
    \hat{H}_{\text{eff}, j} = \sum_{j = 1} ^N -\Delta_{\text{phon}} \hat{b}^\dagger_j \hat{b}_j + \left( \effcoupling_+ \hat{a} \hat{b}_j + \effcoupling_- \hat{a} \hat{b}_j^\dagger + \text{H.c.} \right).
\end{equation}
This effective model is similar to the generalised Dicke model~\cite{Dimer2007pro}, with distinct co- and counter-rotating couplings ($\effcoupling_{\mp}$, respectively).   
However, the effective model in Eq.~\eqref{eq:H_effj} differs from the Dicke model in that it replaces two-level systems by bosons (phonons); this would change the behaviour of such a model above threshold, but does not modify the location of the threshold.  The phase diagram of the generalised Dicke model has been studied in several places~\cite{Keeling2010cdo,Kirton2018sal,Gutierrez-Jauregui2018dqp,Shchadilova2020fff,Stitely2022lac}. We will refer to these works, particularly~\cite{Kirton2018sal}, to interpret features of the phase diagrams we find in the following.

We can derive approximate perturbative forms for the couplings $\lambda_\pm$ and the dressed-cavity detuning $\Deltacdressed$ appearing in this model in the limit where the phonon energy $\omega_{\text{phon}}$ is close to $\Delta\omega$ and all detunings are much smaller than this scale: i.e., for
$\Deltaexdressed, \Delta_c, \Delta_{\text{phon}} \ll \omega_{\text{phon}} \approx \Delta\omega$.
Here we have introduced the dressed exciton detuning $\Deltaexdressed = \Delta_{\text{ex}} + \omega_{\text{phon}}\Delta q^2$, accounting for polaron dressing of the exciton.  That is, $\Deltaexdressed$  corresponds to detuning between the pump and the polaron energy, where the polaron energy is the lowest energy attainable in the one-exciton manifold, corresponding to minimising the Hamiltonian $\omega_{\text{phon}} 
    \left[\hat{b}^\dagger \hat{b}  + \Delta q \left( \hat{b}^\dagger + \hat{b} \right) \hat{X}^\dagger \hat{X} \right]$. 
In the limit where all detunings are much smaller than $\omega_{\text{phon}}$, we find:
\begin{equation}
    \label{eq:effcoupling}
    \effcoupling_{\pm} = \frac{1}{4} \Omega_\pm g_0 \Delta q \left( \frac{1}{\Deltaexdressed \pm \Delta_{\text{phon}}} + \frac{1}{\Deltaexdressed - \Delta_c} \right),
\end{equation}
along with  $\Deltacdressed = \Delta_c + g_0^2 N / \left( \Deltaexdressed - \Delta_c \right)$.  This dressed cavity detuning accounts for hybridisation of photons with excitons.  As well as modifying this detuning, this hybridisation also modifies loss terms, so that cavity loss is increased: $\kappa \to \kappadressed = \kappa + \frac{1}{2} g_0^2 N \Gamma_{\textrm{ex}} / \left( \Deltaexdressed - \Delta_c\right)^2$.  In~\cite{supp}, we provide an alternate numerical approach to give non-perturbative estimates of these dressed energies and decay rates.

While the model above does not provide quantitatively accurate results, it does indicate what conditions might favour strong coupling between phonons and photons, and thus photon-polariton condensation.   From the form of Eq.~\ref{eq:effcoupling}, there are two possible conditions that lead to large values of $\lambda_\pm$, corresponding to either of the two denominators approaching zero. These two cases correspond to the exciton being in resonance with the cavity, or to a vibrational sideband of the exciton being in resonance with one of the pumps.  In both cases, this can lead to a large population of excitons, thus invalidating the adiabatic elimination, and furthermore causing heating due to exciton decay. The ideal scenario is to be very close to the two-photon resonance (i.e., where the Raman process resonantly creates or destroys a phonon), while avoiding the one-photon resonance that leads to large exciton populations. We also note that Eq.~\ref{eq:effcoupling} suggests that the denominators can be made arbitrarily small, leading to arbitrarily large coupling. However, exciton and phonon decay rates introduce cutoffs on these denominators. This consideration motivates the values of the control parameters $\Delta_{\text{ex}}, \Delta_c, \Delta_{\text{phon}}$ used in this work, in that we generally probe the regimes $\left| \Deltaexdressed \right| \gtrsim \Gamma_{\text{ex}}$ and $\left| \Delta_{\text{phon}} \right| \gtrsim \Gamma_{\text{phon}}$. 

\subsection{Superradiant phase transition}
\begin{figure}[t!]
    \centering
    \includegraphics[width=\linewidth]{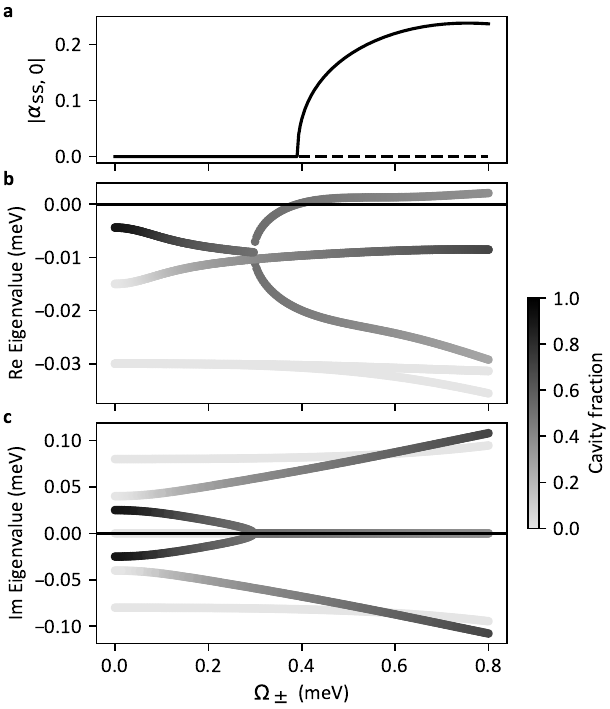}
    \caption{Pitchfork bifurcation leading to a superradiant state. (a) Zero-frequency component of the stationary-state cavity amplitude $\alpha_{\text{ss}, 0}$, showing the emergence of the superradiant phase above the bifurcation at $\Omega_\pm \approx \qty{0.4}{\meV}$. (b,c) Real and imaginary parts of the eigenvalues of the Floquet system linearised about the normal state. Only the 7 nontrivial eigenvalues with the greatest real parts are shown. Shading indicates the fraction of cavity mode in the corresponding eigenvectors, such that darker shading indicates more photonic modes. Parameters are in Table~\ref{tab:parameters}, along with $\Delta_c = \qty{-0.25}{\meV}$.}
    \label{fig:standard_superradiance}
\end{figure}

Considering the full photon-exciton-phonon model of Sec.~\ref{sec:model}, we next discuss the ways in which the system may transition into a superradiant state (i.e., to a phonon-polariton condensate). As discussed above, the $\alpha_{\text{ss},0}=0$ stationary state may go from stable to unstable as indicated by the eigenvalues of the Floquet problem.  Such instability can arise as one increases the driving strength. We discuss here two scenarios of what may happen at this instability, associated with continuous and discontinuous phase transitions.

In the continuous case, the instability of the normal state coincides with the appearance of a new stable stationary state in which $\alpha_{\text{ss}, 0} \neq 0$. Such states correspond to a macroscopic population of the cavity that spontaneously breaks the symmetry of the model.  In the original frame, $\alpha_{\text{ss}, 0}$ corresponds to cavity light at the mean pump frequency $\overline{\omega}$:  The appearance of light at this mean frequency is understood as the signature of a superradiant state~\cite{Dimer2007pro,Kroeze2018sso}.  Stationary states with $\alpha_{\text{ss}, 0} \neq 0$ must come in equal and opposite pairs in the complex plane because the master equation has a (weak~\cite{Buca2012ano,Albert2014sac}) symmetry under the transformation $(\alpha, \hat X) \leftrightarrow (-\alpha, -\hat X)$.
As such, the appearance of nonzero $\alpha_{\text{ss}, 0}$ corresponds to spontaneous symmetry breaking. An illustration of this scenario is shown in Fig.~\ref{fig:standard_superradiance}.  Zooming in near the transition, we see it occurs when a single, real eigenvalue goes from negative to positive, corresponding to a  pitchfork-bifurcation instability~\cite{Strogatz2000nda}.
This corresponds to the standard continuous phase transition to the Hepp-Lieb-Dicke superradiant phase~\cite{Kirton2019itt}.

The critical pumping strength may be estimated using the low-energy effective model introduced above. For effective couplings that are nearly equal, $\left| \effcoupling_{-} - \effcoupling_{+} \right| \ll \effcoupling_{-} + \effcoupling_{+}$,  the transition occurs when~\cite{Kirton2019itt}
\begin{equation}
    \label{eq:sr-threshold}
    \effcoupling_{-} \effcoupling_{+} N = \frac{1}{4} 
    \left(
    \frac{(\Deltacdressed)^2 + (\kappadressed)^2}{\Deltacdressed}
    \right)
    \left(
    \frac{\Delta_{\text{phon}}^2 + \Gamma_{\text{phon}}^2/4}{\Delta_{\text{phon}}} \right).
\end{equation}
The right-hand side of this expression reduces to the more familiar
$\Deltacdressed \Delta_{\text{phon}}/4$ if loss is ignore.  However, including effects of loss is crucial for understanding the minimal threshold pump power attainable, as we discuss further in Sec.~\ref{sec:phase-diagram}.

A notable feature of the low-energy effective model is the replacement of bare cavity loss by the effective loss rate $\kappadressed$.  This effect can be seen directly from Fig.~\ref{fig:standard_superradiance}b.  The intercepts of each line with the vertical axis indicate the linewidths of the effective modes in the absence of pumping, i.e., when $\lambda_\pm = 0$.  One may see that the darkest line, corresponding to the cavity mode, has an intercept much below the bare value of $-\kappa = \qty{-6e-4}{\meV}$. We previously gave the perturbative expression for the dressed linewidth as $\kappadressed = \kappa + \frac{1}{2} g_0^2 N \Gamma_{\textrm{ex}} / \left( \Deltaexdressed - \Delta_c\right)^2$. For the parameters of Fig.~\ref{fig:standard_superradiance}(b), this yields $-\kappadressed \approx \qty{-0.009}{\meV}$, while the intercept in the plot is $\qty{-0.004}{\meV}$. The discrepancy is due to the large value of the parameter that controls the perturbation expansion, $g_0 \sqrt{N} / \left( \Deltaexdressed - \Delta_c \right) \approx 0.5$. One may note that the expansion overestimates the broadening because it disregards the eventual saturation of the dressed cavity linewidth at very large coupling $g_0 \sqrt{N}$:  In the large-coupling limit, the relevant mode becomes an equal superposition of cavity and exciton, with a corresponding linewidth $\kappadressed=\frac{1}{2} \kappa + \frac{1}{4} \Gamma_{\text{ex}}$.

\begin{figure}
    \centering
    \includegraphics[width=\linewidth]{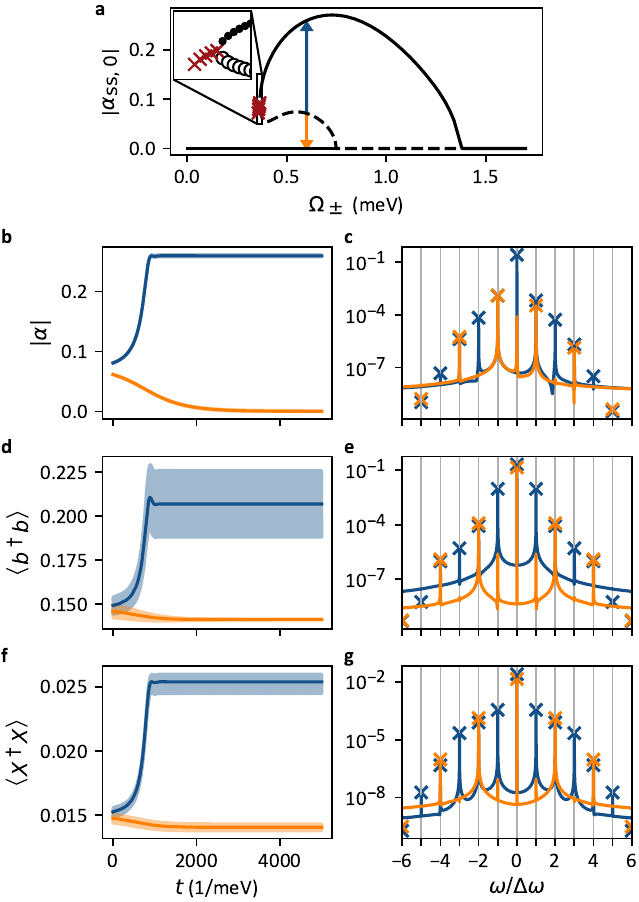}
    \caption{Superradiant bistability. The system is evolved from two initial states labelled with blue and orange throughout. The upper, blue lines approach the stable superradiant phase, while the lower, orange lines approach the stable normal phase, indicating bistability. In (a) the stable stationary states are joined by solid lines, and the unstable states by dashed lines. The red crosses are spurious stationary states found by the solver. Time evolution plots and their Fourier components obtained via both time evolution (lines) and the stationary state solver (crosses) for: (b,c), the cavity amplitude; (d,e), the phonon number; and (f,g), the exciton number. The solid lines on the left are moving averages taken over 2 periods. Fourier components are taken from the time evolution between $4000$ to $\qty[per-mode=symbol]{5000}{\per\meV}$. Even at these late times, the cavity amplitude's zero-frequency component and the phonon and exciton number's $\pm 1\Delta\omega$ frequency components have not fully decayed. Parameter values are in Table~\ref{tab:parameters}, along with $\Delta_c = \qty{-0.24}{\meV}$ and $\Omega_\pm = \qty{0.6}{\meV}$.}
    \label{fig:superradiant_bistability}
\end{figure}

In other parameter regimes, the transition to the superradiant state can be discontinuous.  An example of this is shown in Fig.~\ref{fig:superradiant_bistability}, where a stable superradiant state emerges from a saddle-node bifurcation at nonzero $\alpha_{\text{ss}, 0}$ at a particular driving strength $\Omega$. There is a range of driving strengths above this for which the mean-field theory predicts bistability. At the upper limit of the bistable region, the normal phase goes unstable at a subcritical rather than a supercritical pitchfork bifurcation~\cite{Strogatz2000nda}. This case allows for hysteresis, and it illustrates an unusual first-order superradiant transition~\cite{Keeling2014fsi,Soriente2018dam};  as with other models where mean-field theory supports bistability, beyond-mean-field treatments are expected to predict a unique discontinuous phase transition, determined by the switching rates between the $\alpha_{\text{ss}, 0}=0$ and $\alpha_{\text{ss}, 0}\neq 0$ solutions. Although this transition is discontinuous, we note it still involves spontaneous symmetry breaking to pick the sign of $\alpha_{\text{ss}, 0}$.

In addition to exhibiting a discontinuous transition, Fig.~\ref{fig:superradiant_bistability} also shows another notable feature:  A transition back to the normal state at larger pump strengths.  This can be understood as arising from increasing effective cavity linewidth at strong pumping, due to an increased hybridisation with the exciton; see Ref.~\cite{supp} for further discussion.

\subsection{Lasing transition}

\begin{figure}
    \centering
    \includegraphics[width=\linewidth]{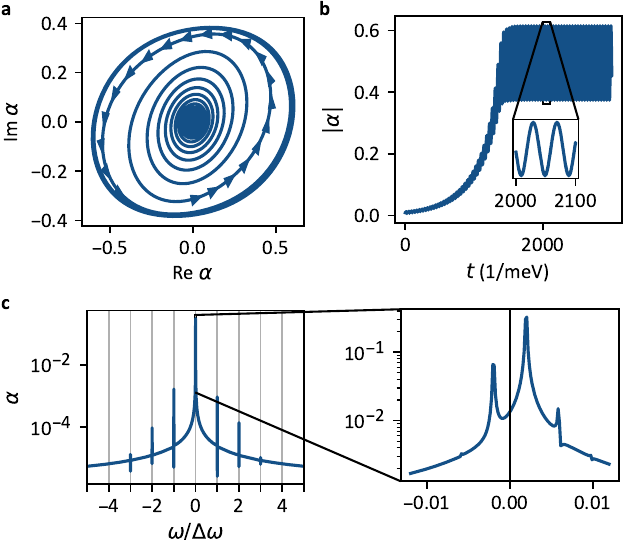}
    \caption{Lasing. The system is initialised close to the unstable stationary state near the vacuum. (a) Real and imaginary parts of the cavity amplitude $\alpha$. (b) Time evolution of $\alpha$, indicating the slow approach to the limit cycle and its associated long time scale. (c) Fourier transform of the late time evolution, from $t = 2000$ to $\qty[per-mode=symbol]{3000}{\per\meV}$.  The inset shows behaviour near zero frequency, indicating that the low-frequency peak is detuned away from zero frequency. Parameter values are in Table~\ref{tab:parameters}, along with $\Delta_c = \qty{-0.08}{\meV}$ and $\Omega_\pm = \qty{1}{\\meV}$.}
    \label{fig:lasing_time_evolution}
\end{figure}

The emergence of a superradiant state is not the only manner in which the normal phase can go unstable.  There is also a scenario where two complex-conjugate eigenvalues simultaneously cross the imaginary axis.  This corresponds to a Hopf bifurcation~\cite{Strogatz2000nda} and is expected to lead to a limit cycle.  Indeed, as seen in Fig.~\ref{fig:lasing_time_evolution}, we find that in this case the cavity amplitude $\alpha$ spirals out from the origin and undergoes large-scale oscillations on a timescale that is much longer than the period of the drive, $1 / \Delta\omega$.  That is, the time dependence of this solution is not commensurate with the drive.  Near to the transition, the frequency of these slow oscillations matches the real part of eigenvalue, as is typical for a Hopf bifurcation. Because the stationary-state solver assumes that the final frequency components are integer multiples of $\Delta \omega$, it cannot find these incommensurate solutions.   As such, we use direct integration of the master equation to check the stability of the limit cycle.

We denote this behaviour the \textit{lasing} phase, in analogy to the phases found in the generalised Dicke model.  After adiabatic elimination of excitons, the generalised Dicke model (as found in Sec.~\ref{sec:leem}) is time independent in the frame of the average pump frequency.  One can distinguish time-independent (superradiant) and time-dependent (lasing) states in such a case. Reference~\cite{Kirton2018sal} showed that there are two different parameter regimes where the generalised Dicke model exhibits lasing.  The first is the typical lasing regime, where an incoherent drive leads to inversion of the emitters, causing gain for the cavity mode.   However, the model of Sec.~\ref{sec:leem} has no incoherent drive, only decay, and the emitters in Eq.~\ref{eq:effcoupling} are bosons, not two-level systems.  As such, inversion of emitters is not relevant for the case we consider here.  The second lasing state has a different character.  This state is found either when $\effcoupling_+ \gg \effcoupling_-$, or when the sum of the cavity and emitter frequencies approach zero.  In either of these cases, the dominant matter-light coupling is one that creates a photon and \emph{excites} an emitter, the counter-rotating coupling.  This then leads to a state that has been referred to as ``counter-lasing''~\cite{Shchadilova2020fff,Stitely2022lac}.

In the case shown in Fig.~\ref{fig:lasing_time_evolution} (and also discussed further below), our time-dependent phase can be ascribed to counter-lasing driven by near-resonance of the process that creates a photon and excites an emitter; see, e.g., Fig.~8 of Ref.~\cite{Kirton2018sal}.  Such resonance would occur if $\Deltacdressed=-\Delta_{\text{phon}}$.  Even though the bare detuning $\Delta_c$ is negative in Fig.~\ref{fig:lasing_time_evolution}, corresponding to red detuning, the shift due to the exciton-cavity coupling can push $\Deltacdressed$ to become positive, corresponding to effective blue detuning. In principle, counter-lasing may also occur in our model due to $\effcoupling_+ \gg \effcoupling_-$, but this requires larger $\Delta_{\text{phon}}$ and takes the system out of the small-detuning regime used to derive $\effcoupling_\pm$.

\subsection{Phase Diagram}
\label{sec:phase-diagram}

\begin{figure}
    \centering
    \includegraphics[width=\linewidth]{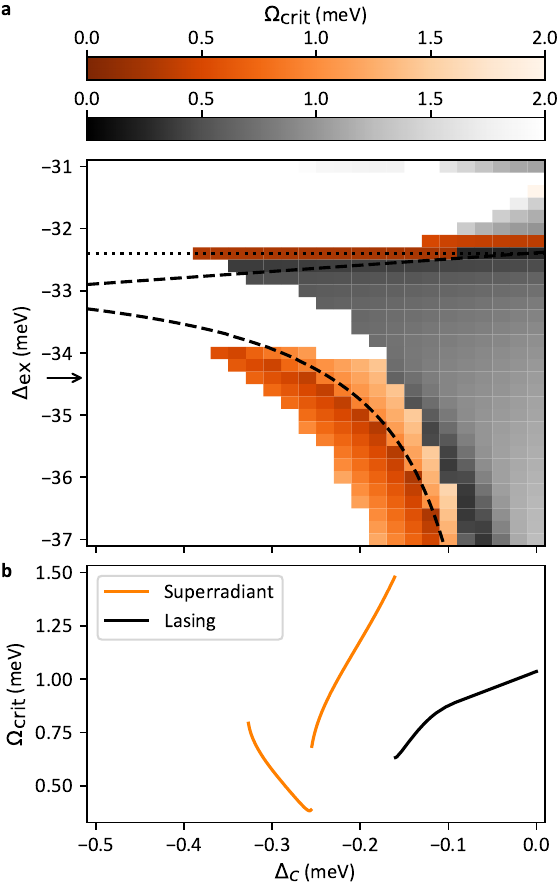}
    \caption{Phase diagram indicating the minimum pump strength at which the normal phase goes unstable. 
    The two colours indicate whether one eigenvalue goes unstable (orange) corresponding to the superradiant transition, or two eigenvalues go unstable (black) corresponding to lasing. Parameters are in Table~\ref{tab:parameters}, along with $\Omega_- = \Omega_+$. (a) The dotted line indicates where $\Deltaexdressed=0$, i.e., where $\Delta_{\text{ex}}$ crosses the polaron energy $-\omega_{\text{phon}} \Delta q^2 = \qty{-32.4}{\meV}$. The dashed lines separate the regions where the cavity eigenvalue at zero pump, i.e., at $\Omega_\pm = 0$, has an imaginary part that is positive (lower right region) versus one that is negative (lower left and upper regions). These regions correspond to the cavity being effectively blue-detuned and red-detuned, respectively. The arrow indicates the value of $\Delta_{\text{ex}}$ for the slice in (b). (b) Horizontal slice of the phase diagram of (a) at the location indicated by the arrow, $\Delta_{\text{ex}} = \qty{-34.4}{\meV}$.}
    \label{fig:phase_diagram}
\end{figure}

Having discussed the forms of instability that can arise, we now discuss the phase diagram, indicating which forms of instability occur for certain parameters. A key practical question is what critical pump strength is required to reach either the superradiant or lasing instability. Figure~\ref{fig:phase_diagram} shows this diagram, mapping out the critical pump strength at which the normal phase goes unstable due to a single eigenvalue (superradiant state) or a pair of complex-conjugate eigenvalues (lasing).  Dark shading indicates where a transition can occur at low pump power.  White regions indicate parameters for which no instability of the normal state is found.

Experimentally exploring this full phase diagram requires tuning both the mean pump frequency $\overline{\omega}$ and the cavity frequency.  Changing only the mean pump frequency corresponds to diagonal traces through this figure from lower left to upper right. 
The superradiant and lasing phases appear continuously connected in the space of $\Delta_c$ and $\Delta_{\text{ex}}$ in some regimes, but for other regimes they are separated. Details of this structure can be seen in the cross section in Fig.~\ref{fig:phase_diagram}(b). A jump in the critical pump strength within the superradiant phase is also exhibited.  This is associated with switching between the continuous and discontinuous phase transitions. To see how this scenario evolves in more detail, Sec.~\ref{sec:eigenvalue-evolution} of Ref.~\cite{supp} shows how eigenvalues coming close to but not crossing the imaginary axis can result in jumps in the transition pumping strength.

A notable feature of this figure is the change in behaviour as the exciton detuning $\Delta_{\text{ex}}$ is tuned through $-\omega_{\text{phon}} \Delta q^2$ as indicated by the dotted line.  This corresponds to  $\Deltaexdressed$ crossing zero, or equivalently, to tuning the mean pump frequency through the polaron energy. At this detuning we see a narrow sliver of superradiant state. One may note that the low-energy effective model does not apply under this condition because resonance results in stationary states that have high-exciton population even at low pump strength. A second consequence of small $\Deltaexdressed$ is a very large shift to the dressed photon frequency;  i.e., $\Deltacdressed$ is large and positive just below $\Deltaexdressed=0$, and large and negative just above this line.  This in turn explains why the superradiant state ceases for $\Delta_{\text{ex}}>-\omega_{\text{phon}} \Delta q^2$.

A second important feature occurs when $\Deltacdressed$ changes sign, as shown by the dashed curves in Fig.~\ref{fig:phase_diagram}.  As discussed in the previous section, counter-lasing occurs in general for positive $\Deltacdressed$, and in particular when $\Deltacdressed \simeq -\Delta_{\text{phon}}$.  In contrast, the superradiant state occurs for negative $\Deltacdressed$. Since Fig.~\ref{fig:phase_diagram} is plotted at fixed $\Delta\omega$ and thus fixed $\Delta_{\text{phon}}$, the switch from the superradiant phase to counter-lasing is driven by changing $\Deltacdressed$.

As can be understood from the low-energy effective model (i.e., Eq.~\eqref{eq:effcoupling} and Eq.~\eqref{eq:sr-threshold}), the points of particularly low threshold are close to points where the transition ceases to exist.  This corresponds to the standard observation that with dissipation, the lowest threshold occurs when $\Deltacdressed=\kappadressed$, while the transition vanishes at $\Deltacdressed=0$.  In Fig.~\ref{fig:phase_diagram}(a), the dashed line corresponding to $\Deltacdressed=0$ does not exactly match the point where the superradiant transition ceases; this is because the estimate of $\Deltacdressed$ used here neglects effects of the AC Stark shift of the exciton due to the finite pump strength $\Omega$.  In contrast, calculations in Ref.~\cite{supp} performed for the low-energy effective model (i.e., neglecting this AC Stark shift) do show the expected agreement.

\section{Discussion}

In the results above we have shown that, considering a standard model of Holstein coupling between excitons and phonons, one can realise a Raman-driving scheme to couple Raman-active phonons to light.  One can thus realise a phonon-polariton condensate. Our results are written in terms of the driving strength $\Omega_\pm$ appearing in Eq.~\ref{eq:Horig}.  Here we discuss the practicality of attaining these values in terms of required laser power.  We also discuss design principles for choices of materials as well as open questions arising from this work.

To estimate the laser power required to reach the superradiant transition, we make several simplifications to obtain a closed-form expression. First, we set the detunings $\Delta_{\text{ex}}^D$ and $\Delta_{\text{phon}}$ to half the exciton and phonon linewidths, respectively. Second, we assume we work in the regime of large exciton-cavity coupling;  in this regime, as discussed above, the dressed-cavity linewidth $\kappa^D \simeq \Gamma_{\text{ex}}/4$, a linear scaling with exciton linewidth.  (We discuss the other possible regimes in Ref.~\cite{supp}). Third, we set the denominators in the effective couplings Eq.~\eqref{eq:effcoupling} to $\Gamma_{\text{ex}}$ because $\Delta_{\text{ex}}^D$ is the largest energy scale in the denominators and is assumed to be dressed by exciton decay of an amount between $\frac{1}{2} \Gamma_{\text{ex}}$ to $\Gamma_{\text{ex}}$. Finally, we choose equal pumping, so $\Omega_\pm = \Omega$. With these choices, the threshold condition becomes $\left( \Omega_{\text{crit}} g_0 \Delta q \right)^2 N = \frac{1}{2} \Gamma_{\text{ex}}^3 \Gamma_{\text{phon}}$. 

To convert this from a critical value of $\Omega$ to a critical laser power, we must relate the light-matter coupling parameters in our model to electric field strengths of the cavity and pump light.
The exciton-light couplings can be written as $\Omega \sqrt{N} = e_{\text{ex}} \sqrt{A} E_p$ and $g_0 \sqrt{N} = e_{\text{ex}} \sqrt{A} E_c$ for pump and cavity electric fields $E_{p, c}$, material area $A$, and exciton matrix element $e_{\text{ex}}$; see \cite{Soh2018ono} and~\cite{supp} for details. We write the number of emitters as $N = n A$ for emitter density $n$. Assuming the pump is a Gaussian beam impinging onto the material area, its power is $P = \frac{1}{4} \varepsilon_0 E_p^2 A c$. The threshold pump power is then:
\begin{equation}
    \label{eq:threshold-power}
    P_{\text{crit}} = \frac{\Gamma_{\text{ex}}^3 \Gamma_{\text{phon}} n \varepsilon_0 c}{8 e_{\text{ex}}^4 \Delta q^2 E_c^2}.
\end{equation}
We see from this expression that the critical pump power depends separately on the density of emitters $n$ and the matrix element $e_{\text{ex}}$, and not just upon the combination $e_{\text{ex}}/\sqrt{n}$ that controls the coupling strength $g_0$.
Since the materials we wish to consider do not actually consist of a set of localised emitters, a choice must be made to give separate estimates of $n$ and $e_{\text{ex}}$. In the following we will set the emitter density to the inverse unit cell area---alternate choices are the Mott density and inverse exciton area \cite{Keeling2007cci}.

We take \ce{MoSe2} as our model material, so we set $\Gamma_{\text{ex}} = \qty{1}{\meV}$, $\Gamma_{\text{phon}} = \qty{0.05}{\meV}$, $e_{\text{ex}} = 0.23 e$, $\Delta q^2 = 1$, and $n^{-1} = \sqrt{3} a_0^2 / 2$ for a lattice spacing of $a_0 = \qty{3.3}{\angstrom}$. We set the cavity waist to $w_0 = \qty{30}{\micro\metre}$ and quantise its electric field over cavity length $L = \qty{1}{\centi\metre}$ with a Gaussian mode profile. This yields a threshold pump power of $P_{\text{crit}} = \qty{30.}{\W}$. While this is too great to be immediately feasible, we will remark later on how a multimode cavity may bring this down to a practical regime. Note that the dependence of the critical power on $e_{\text{ex}}^4$ and $\Gamma_{\text{ex}}^3$ mean small changes in estimates of these parameters have large effects on the critical power.

As mentioned before, the model we discuss is not specific to TMDs and can be applied to other materials.  This prompts the question of which material properties would be best-suited to realising phonon-polariton condensation.  Based on the results above, an estimate of this can be found by considering how to maximise the effective coupling defined in Eq.~\eqref{eq:effcoupling} or minimise the threshold power of Eq.~\eqref{eq:threshold-power}. These expressions highlight the beneficial effects of the exciton matrix element and the Huang-Rhys parameter $S = \Delta q^2$ defining the exciton-phonon coupling, as well as the adverse effect of the exciton linewidth.

To conclude this discussion, we briefly mention some topics that could be addressed in future work.

\emph{Multimode cavities.}
This paper considers a single-mode cavity for simplicity, but experiments using a degenerate confocal cavity~\cite{Kollar2015aac} have been realised for ultracold atoms.  In such confocal cavities, one can obtain a field enhancement~\cite{Vaidya2018tpa,Kroeze2023hcu} by constructive interference between different degenerate modes building a synthetic cavity mode localised to have perfect overlap with the cloud of atoms.  Such an approach can also be used in this solid-state context to maximise overlap with whichever is smaller, the sample or the pump spot.  Such an approach thus allows one to further enhance the cavity-exciton coupling. In particular, a multimode cavity may reduce the threshold laser power required. To continue the calculation from before, if one considers a pump beam waist of $w_p = \qty{5}{\micro\metre}$, then the confocal enhancement would be $\left( w_0 / w_p \right)^2 = 36$, reducing the critical pump threshold to below $\qty{1}{\watt}$.

\emph{Temperature.}
In our calculations above, we implicitly assumed a low-temperature state, in which there is no population of phonons in the initial state.  Such an approach is reasonable when considering experiments at cryogenic temperatures, i.e., when temperature is much smaller than the phonon energy.  Extending the calculations here to consider materials at higher temperature (e.g., room temperature) is an open question.  In particular, at higher temperatures it may be important to distinguish between the exciton and phonon linewidths due to decay versus dephasing.  As has been explored in simple models~\cite{Kirton2018sal}, dephasing and decay have different effects on the threshold for condensation.

\emph{Effects on electronic states.}
A key question for future work is to assess the effect of phonon polariton condensation on electronic and transport properties, following methods such as those in Refs.~\cite{Sentef2018cqp,Mankowsky2014nld,schlawin2022cqm}. A related question is to explore whether other states can be realised by the Raman pumping scheme in this work, such as more complex limit cycles or chaotic phases.  

\section{Methods}
\subsection{Transformation to mean-field Hamiltonian}

To transform the original Hamiltonian, Eq.~\eqref{eq:Horig}, into the frame rotating at $\overline{\omega}$, we use the unitary transform:
\begin{equation}
    \label{eq:Hrot}
    \hat{U}_{\text{rot}} = \exp \left[ i  \overline{\omega} t \left( \hat{a}^\dagger \hat{a} + \sum_{j = 1} ^N \hat{X}_j^\dagger \hat{X}_j \right)\right].
\end{equation}
To derive the model in Eqs.~\eqref{eq:mf1}--\eqref{eq:mfHmat}, we then disregard terms rotating at $2\overline{\omega}$ and $\overline{\omega} + \omega_\pm$.  We then make a mean-field approximation, equivalent to assuming the density matrix factorises into terms for the cavity and for each site.  With this approximation, the equation of motion for the cavity becomes that of a coherently driven damped harmonic oscillator, and so its behaviour is captured by considering the evolution of the coherent state amplitude:
\begin{equation}
    \alpha \equiv \frac{\expval{\hat{a}}}{\sqrt{N}}.
\end{equation}
For the matter component, we assume all sites are equivalent, leading to the site-independent model in Eq.~\eqref{eq:mfHmat}.

\subsection{Fourier analysis and linear stability}

The Fourier series ansatz for stationary states transforms the system of nonlinear differential equations for $(\alpha, \rho)$ into a system of nonlinear algebraic equations for $\left( \alpha_{\text{ss}, n}, \rho_{\text{ss}, n} \right)$.   All nonlinearity here arises from the dependence on $\alpha_{\text{ss},n}$.  As such, at fixed $\alpha_{\text{ss}, n}$, the problem  simplifies into a linear homogeneous system for $\rho_{\text{ss}, n}$. The trace constraint $\text{Tr}\left( \rho_{\text{ss}, n \neq 0} \right) = 0$ is already enforced by the system, but we must manually add the normalisation constraint $\text{Tr}\left( \rho_{\text{ss}, 0} \right) = 1$ to one of the rows. We then solve this exactly determined system for $\rho_{\text{ss}, n}$ via sparse LU factorisation, and we feed the result back into the objective function for $\alpha_{\text{ss}, n}$. This reduces the size of the nonlinear system from the full space of $\left( \alpha_{\text{ss}, n}, \rho_{\text{ss}, n} \right)$ to the $2M + 1$ complex numbers $\alpha_{\text{ss}, n}$, where $M$ is the maximum $|n|$ retained in the sum; i.e., $M$ is the cutoff on Fourier components. We also find the Jacobian of the objective function of $\alpha_{\text{ss}, n}$ to accelerate the nonlinear root-finding. The Fourier cutoff $M$ is typically set to 4, but it is occasionally set to 6 for detailed comparison to time evolution, as indicated in Fig.~\ref{fig:superradiant_bistability} and Table \ref{tab:parameters}. We use the default settings in both the sparse LU factorisation and the root-finding using the Levenberg-Marquardt algorithm~\cite{Levenberg1944amf,Marquardt1963aaf}, as implemented in SciPy~\cite{Virtanen2020s1f} following MINPACK~\cite{More1980ugf}.

The stationary-state solver returns a solution $\left( \alpha_{\text{ss}}(t), \rho_{\text{ss}}(t) \right)$ with period $T = 2\pi / \Delta\omega$. To determine linear stability, we assume a small fluctuation about this stationary state: $\alpha(t) = \alpha_{\text{ss}}(t) + \delta\alpha(t)$ and $\rho(t) = \rho_{\text{ss}}(t) + \delta\rho(t)$, from which we obtain a system of differential equations for $\left( \delta\alpha(t), \delta\rho(t) \right)$ that is linear but with explicit, periodic time dependence. Writing this system as $\partial_t \vec{x} = \hat{A}(t) \vec{x}$ for $\vec{x} = (\delta\alpha, \delta\rho)$ and $T$-periodic $\hat{A}(t)$, we integrate to find the one-period evolution operator, also termed the monodromy matrix~\cite{Glendinning1994sia}:
\begin{equation}
    \label{eq:monodromy}
    \hat{\mathcal{M}} = \text{T}_\leftarrow \exp\left[ \int_0 ^T \hat{A}(t)\, dt \right],
\end{equation}
where $\text{T}_\leftarrow$ is the time-ordering symbol. We perform this integral by exponentiating and multiplying over discrete time steps of size $T / 100$; we find this faster and more stable than any more sophisticated methods. The slow step is the subsequent diagonalisation of $\hat{\mathcal{M}}$, yielding eigenvalues $e^{\mu_k T}$ for Floquet multipliers $\mu_k$. As mentioned before, these $\mu_k$ are called eigenvalues in this manuscript.

Direct time evolution of the system is performed using Netlib's zvode, as implemented in SciPy \cite{Virtanen2020s1f}. We use the implicit Adams method at order 12, with rtol and atol parameters of $10^{-9}$ and $10^{-12}$, respectively, and with a maximum number of steps of 1000.  

Table~\ref{tab:parameters} provides a summary of parameter values used in our numerical calculations. Throughout, QuTiP~\cite{Johansson2013q2a} is used to create sparse matrices and to interpret density matrices, though the mean-field setup is not directly amenable to QuTiP methods.

\begin{table}[b!] 
\centering
\caption{Full parameter list including values used in simulations.}
\begin{tblr}{width=0.48\textwidth,colspec={Q[c,m]X[c,m]Q[c,m]}} \hline \hline
Symbol & Meaning & Value (meV) \\ \hline
$\omega_{\text{phon}}$ & Phonon energy & 40 \\
$\Delta_{\text{ex}}$ & Exciton detuning $\overline{\omega} - \omega_{\text{ex}}$ & $-34.5$, varies \\
$\Delta_c$ & Cavity detuning $\overline{\omega} - \omega_c$ & varies \\
$\Delta \omega$ & Detuning from the mean pump to either pump $\left| \overline{\omega} - \omega_\pm  \right|$ & $39.96$ \\
$\Delta_{\text{phon}}$ & Detuning from $\Delta\omega$ to the phonon energy $\Delta\omega - \omega_{\text{phon}}$ & $-0.04$ \\ \hline
$\Omega_\pm$ & Exciton-pump Rabi frequency & varies \\
$g_0 \sqrt{N}$ & Mean-field exciton-cavity coupling & $1$ \\ \hline
$\Gamma_{\text{ex}}$ & Exciton number decay rate & 0.06 \\
$\Gamma_{\text{phon}}$ & Phonon number decay rate & 0.03 \\
$\kappa$ & Mean-field cavity amplitude decay rate & $6 \times 10^{-4}$ \\ \hline \hline
Symbol & Meaning & Value \\ \hline
$\Delta q$ & Exciton-phonon coupling scaled by $\omega_{\text{phon}}$; Huang-Rhys parameter $S = \Delta q^2$ & 0.9 \\
$\text{n}_{\text{ex}}$ & Maximum exciton number & 3 \\
$\text{n}_{\text{phon}}$ & Maximum phonon number & 7 \\
$M$ & Fourier cutoff & 4, 6 \\ \hline \hline
\end{tblr} \label{tab:parameters}
\end{table}

\section{Acknowledgements}
We acknowledge helpful discussions with T.~Heinz, A.~J.~Daley, H.~S.~Hiller, D.~Lao, R.~M.~Kroeze, B.~P.~Marsh, and H.~S.~Hunt. JK and BLL acknowledge funding from the Gordon and Betty Moore Foundation. Large-scale simulations were performed on the Sherlock cluster, provided by Stanford University and the Stanford Research Computing Center.


%

\clearpage
\onecolumngrid
\section*{Supplementary Information}

\renewcommand{\theequation}{S\arabic{equation}}
\renewcommand{\thefigure}{S\arabic{figure}}
\setcounter{figure}{0}
\setcounter{equation}{0} 
\setcounter{section}{0} 

\section{Derivation of the Low-Energy Effective Hamiltonian}

This section derives the effective photon-phonon model and couplings given in Eq.~\eqref{eq:effcoupling}. We first perform the rotating-frame transformation generated by Eq.~\eqref{eq:Hrot} and disregard counter-rotating terms. We then apply the Lang-Firsov polaron transform~\cite{Lang1963kto} to the result:
\begin{equation}
    \hat{H} \to \hat{H}_{\text{L-F}} = e^{\hat{S}_{\text{L-F}}} \hat{H} e^{-\hat{S}_{\text{L-F}}};
    \qquad
    \hat{S}_{\text{L-F}} = \sum_{j = 1} ^N \Delta q \left( \hat{b}^\dagger_j - \hat{b}_j \right) \hat{X}^\dagger_j \hat{X}_j.
\end{equation}
We divide the resulting Hamiltonian into terms that maintain and change the exciton number: $\hat{H}_{\text{L-F}} = \hat{H}_0 + \hat{H}_1$, with
\begin{align}
    \hat{H}_0 &= -\Delta_c \hat{a}^\dagger \hat{a} + \sum_{j = 1} ^N \left[ \left( -\Delta_{\text{ex}} - \omega_{\text{phon}} \Delta q^2 \right) \hat{X}^\dagger_j \hat{X}_j + \omega_{\text{phon}} \hat{b}^\dagger_j \hat{b}_j - \omega_{\text{phon}} \Delta q^2 \hat{X}^\dagger_j \hat{X}^\dagger_j \hat{X}_j \hat{X}_j \right]; \\
    \hat{H}_1 &= \sum_{j = 1} ^N \left[ \left( \Omega_+ \frac{e^{-i\omega_+ t}}{2} + \Omega_- \frac{e^{i \omega_- t}}{2} + g_0 \hat{a} \right) \hat{X}^\dagger_j e^{\Delta q \left( \hat{b}^\dagger_j - \hat{b}_j \right)} + \text{H.c.} \right].
\end{align}
We define the phonon-dressed exciton detuning $-\Deltaexdressed \equiv -\Delta_{\text{ex}} - \omega_{\text{phon}} \Delta q^2$.  In the following, we will ignore the anharmonic exciton term $X^\dagger X^\dagger X X$. 
 
Next we use the Schrieffer-Wolff transformation~\cite{Schrieffer1966rbt,Bhaseen2012don} to isolate the zero-exciton subspace, which requires finding a suitable generator $\hat{S}_{\text{S-W}}$ such that $\left[ \hat{S}_{\text{S-W}}, \hat{H}_0 \right] = -\hat{H}_1$. It will be helpful to temporarily use the symbols $\hat{\Omega}^k$ and $\Delta_{\text{ex}}^k$ in describing the terms coupling to the exciton, where $k$ runs over $+$, $-$, and $c$:
\begin{alignat}{3}
    &\hat{\Omega}^+ = \Omega_+ \frac{e^{-i \Delta\omega t}}{2}; \quad &&\hat{\Omega}^- = \Omega_- \frac{e^{+i \Delta\omega t}}{2}; \quad &&\hat{\Omega}^c = g_0 \hat{a}; \\
    &\Delta_{\text{ex}}^+ = \Deltaexdressed + \Delta\omega; \quad &&\Delta_{\text{ex}}^- = \Deltaexdressed - \Delta\omega; \quad &&\Delta_{\text{ex}}^c = \Deltaexdressed - \Delta_c.
\end{alignat}
With these, we may state the Schrieffer-Wolff generator $\hat{S}_{\text{S-W}}$. 
We note that formally this is an expansion in $\Delta q$, yet in the main text $\Delta q = 0.9$, which is expected to be outside the range of a perturbative expansion.  Nonetheless, the results of the perturbative expansion are illustrative for a qualitative understanding, even if they are not quantitatively accurate. The generator is:
\begin{equation}
    \hat{S}_{\text{S-W}} = - \sum_{\substack{j = 1..N \\ k \in \{+, -, c\}}} \left[ \hat{\Omega}^k \left( \frac{1}{\Delta_{\text{ex}}^k} + \frac{\Delta q}{\Delta_{\text{ex}}^k - \omega_{\text{phon}}} \hat{b}^\dagger_j - \frac{\Delta q}{\Delta_{\text{ex}}^k + \omega_{\text{phon}}} \hat{b}_j \right) - \text{H.c.} \right] + \mathcal{O}(\Delta q)^2.
\end{equation}
Then, to order $\Delta q^1$, the zero-exciton subspace of the resulting Hamiltonian $\hat{H}_{\text{S-W}} = e^{\hat{S}_\text{S-W}} \hat{H}_{\text{L-F}} e^{-\hat{S}_{\text{S-W}}}$ is:
\begin{equation}
    \hat{H}_{\text{S-W}} = \hat{H}_0 + \frac{1}{2} \sum_{\substack{j = 1..N \\ k, k^\prime \in \{+, -, c\}}} \left[ \left( \hat{\Omega}^{k^\prime} \right)^\dagger \hat{\Omega}^k \left( 1 - \Delta q \hat{b}^\dagger_j + \Delta q \hat{b}_j \right) \left( \frac{1}{\Delta_{\text{ex}}^k} + \frac{\Delta q}{\Delta_{\text{ex}}^k - \omega_{\text{phon}}} \hat{b}^\dagger_j - \frac{\Delta q}{\Delta_{\text{ex}}^k + \omega_{\text{phon}}} \hat{b}_j \right) + \text{H.c.} \right].
\end{equation}
Already, we see that a process that creates or annihilates a phonon has two routes with opposite signs and whose denominators differ by $\omega_{\text{phon}}$. 

The final step is to eliminate the explicit time dependence in the one-phonon terms by rotating the phonons at $\Delta\omega$ via the unitary $\hat{U}_{\text{rot,\,phon}} = \exp\left( i \Delta\omega t \sum_{j = 1} ^N  \hat{b}^\dagger_j \hat{b}_j \right)$.
This transform changes the phonon energy from $\omega_{\text{phon}}$ to $\omega_{\text{phon}} - \Delta\omega \equiv -\Delta_{\text{phon}}$.
After this transform, the time-independent photon-phonon coupling terms have the form:
\begin{equation}
\label{eq:Hab}
\begin{split}
    \hat{H}_{\text{a-b}} &= \frac{g_0 \omega_{\text{phon}} \Delta q}{4} \sum_{j = 1} ^N \left\{ \Omega_+ \left[ \frac{1}{ \left( \Deltaexdressed + \Delta\omega \right) \left( \Deltaexdressed + \Delta_{\text{phon}} \right)} + \frac{1}{ \left( \Deltaexdressed - \Delta_c \right) \left(\Deltaexdressed - \Delta_c + \omega_{\text{phon}} \right)} \right] \left( \hat{a} \hat{b}_j + \text{H.c.} \right) \right. \\
    &\phantom{= \frac{g \omega_{\text{phon}} \Delta q}{4} \sum_{j = 1} ^N \Big\{ }\ \left. +\, \Omega_- \left[ \frac{1}{ \left( \Deltaexdressed - \Delta\omega \right) \left( \Deltaexdressed - \Delta_{\text{phon}} \right)} + \frac{1}{ \left( \Deltaexdressed - \Delta_c \right) \left( \Deltaexdressed - \Delta_c - \omega_{\text{phon}} \right)} \right] \left( \hat{a} \hat{b}^\dagger_j + \text{H.c.} \right) \right\}.
\end{split}
\end{equation}
The denominators in this expression are each a product of two values that differ by an amount $\Delta\omega \approx \omega_{\text{phon}}$.
In the main text we discussed the limit where the detunings are optimised so that we take $\Deltaexdressed, \Delta_c, \Delta_{\text{phon}} \ll \Delta\omega \approx \omega_{\text{phon}}$.
In this limit, we may assume the second factor in the denominator is approximately $\omega_{\text{phon}}$, and thus we can cancel this factor with the prefactor $\omega_{\text{phon}}$ appearing in Eq.~\eqref{eq:Hab}.  This then gives the effective couplings $\effcoupling_\pm$ as stated in Eq.~\eqref{eq:effcoupling}.

The $\left( \hat{\Omega}^{k^\prime} \right)^\dagger \hat{\Omega}^k$ terms of $\hat{H}_{\text{S-W}}$ also include terms proportional to $\hat{a}^\dagger \hat{a}$.  These terms mean there is a shift of the cavity frequency due to coupling to excitons.  This results in a dressed-cavity detuning $-\Deltacdressed = -\Delta_c + g_0^2 N / \left( \Deltaexdressed - \Delta_c \right)$, as introduced in the main text.  
There are also displacements to $\hat{a}$ and $\hat{b}_j$ in $\hat{H}_{\text{S-W}}$, but those do not affect the photon-phonon coupling. 

Combining all the above, the terms of the effective Hamiltonian are:
\begin{equation}
    \label{eq:Heff_supp}
    \hat{H}_{\text{eff}} = -\Deltacdressed \hat{a}^\dagger \hat{a} + \sum_{j = 1} ^N \left[ -\Delta_{\text{phon}} \hat{b}_j^\dagger \hat{b}_j + \effcoupling_+ \left( \hat{a} \hat{b}_j + \text{H.c.} \right) + \effcoupling_- \left( \hat{a} \hat{b}_j^\dagger + \text{H.c.} \right) \right],
\end{equation}
as reported in Eq.~\eqref{eq:H_effj}. When $\left| \effcoupling_+ - \effcoupling_- \right| \ll \effcoupling_+ + \effcoupling_-$, the condition for instability of the normal phase is given in Eq.~\eqref{eq:sr-threshold}.

\begin{figure}
    \centering
    \includegraphics[scale=1]{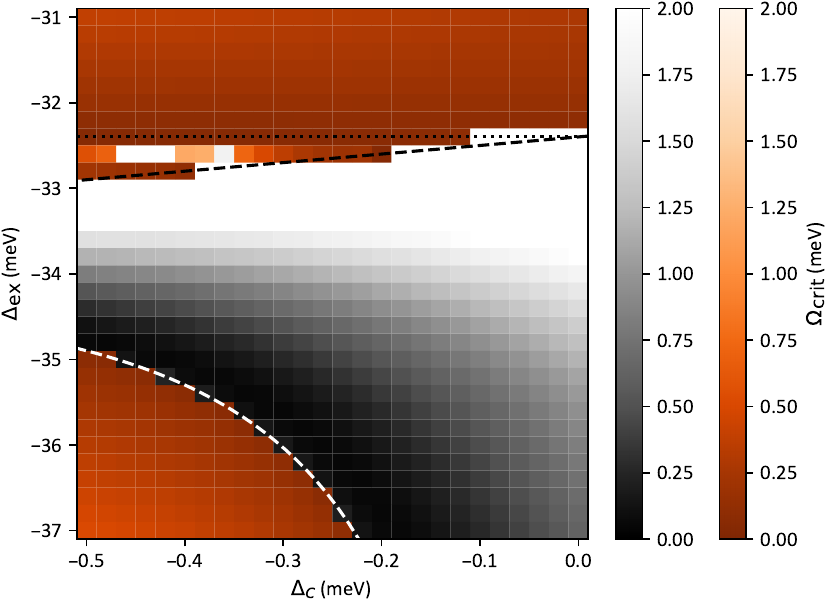}
    \caption{Phase diagram indicating the minimum pump strength at which the normal phase goes unstable in the low-energy effective model. 
    As in Fig.~\ref{fig:phase_diagram}, the two colours indicate whether one eigenvalue goes unstable (orange) corresponding to the superradiant transition, or two eigenvalues go unstable (black) corresponding to lasing. Parameters are in Table~\ref{tab:parameters}, along with $\Omega_- = \Omega_+$. The dotted and dashed lines have the same meaning as in the main text.  That is, the dotted line indicates where $\Deltaexdressed = 0$. The dashed line is defined by taking the eigenvalue which which at zero pump strength corresponds to the cavity mode, and determining when it has an imaginary part (i.e.,~oscillation frequency) that is positive (lower right region) or that is negative (lower left and upper regions).}
    \label{fig:eff_phase_diagram}
\end{figure}

A full treatment of this approximate model must also consider dissipative terms.  A rigorous approach to this would require that one derive the effective master equation within the restricted subspace.  A first approximation to this is to replace all factors of the form $1 / \left( \Deltaexdressed + \bullet \right)$ with $\operatorname{Re}{\left[ 1 / \left( \Deltaexdressed - i \Gamma_{\text{ex}}/2 + \bullet \right) \right]}$. Including such damping regularises the divergence that might otherwise arise from such denominators.

The mean-field equations of motion for $\alpha \equiv \expval{\hat{a}} / \sqrt{N}$ and $\beta \equiv \expval{\hat{b}_j}$ evolving under Eq.~\eqref{eq:Heff_supp} always admit the normal-phase steady-state solution $\alpha = \beta = 0$. Since $\hat{H}_{\text{eff}}$ is time-independent this state is truly steady rather than just stationary as in the main text. Linearising about this fixed point of the dynamics yields a $4 \times 4$ system for fluctuations of $\alpha, \alpha^*, \beta, \beta^*$, and the eigenvalues of this system determine the stability of the steady state. As in the main text, one can solve for when either one or two eigenvalues gain a positive real part, corresponding to instability to the superradiant or lasing state, respectively. The resulting phase diagram (including effects of dissipation as discussed in the previous paragraph) is presented in Fig.~\ref{fig:eff_phase_diagram}. This is qualitatively similar to the one found with much greater computational effort in the main text, but there are noticeable quantitative differences in the superradiant-to-lasing crossover as well as the required pumping power.  Notably in this figure the point where $\Deltacdressed=0$ does indeed correspond to the switch between the superradiant and lasing states.

\section{Eigenvalue evolution in different regimes of cavity detuning}
\label{sec:eigenvalue-evolution}

\begin{figure}
    \centering
    \includegraphics[width=\textwidth]{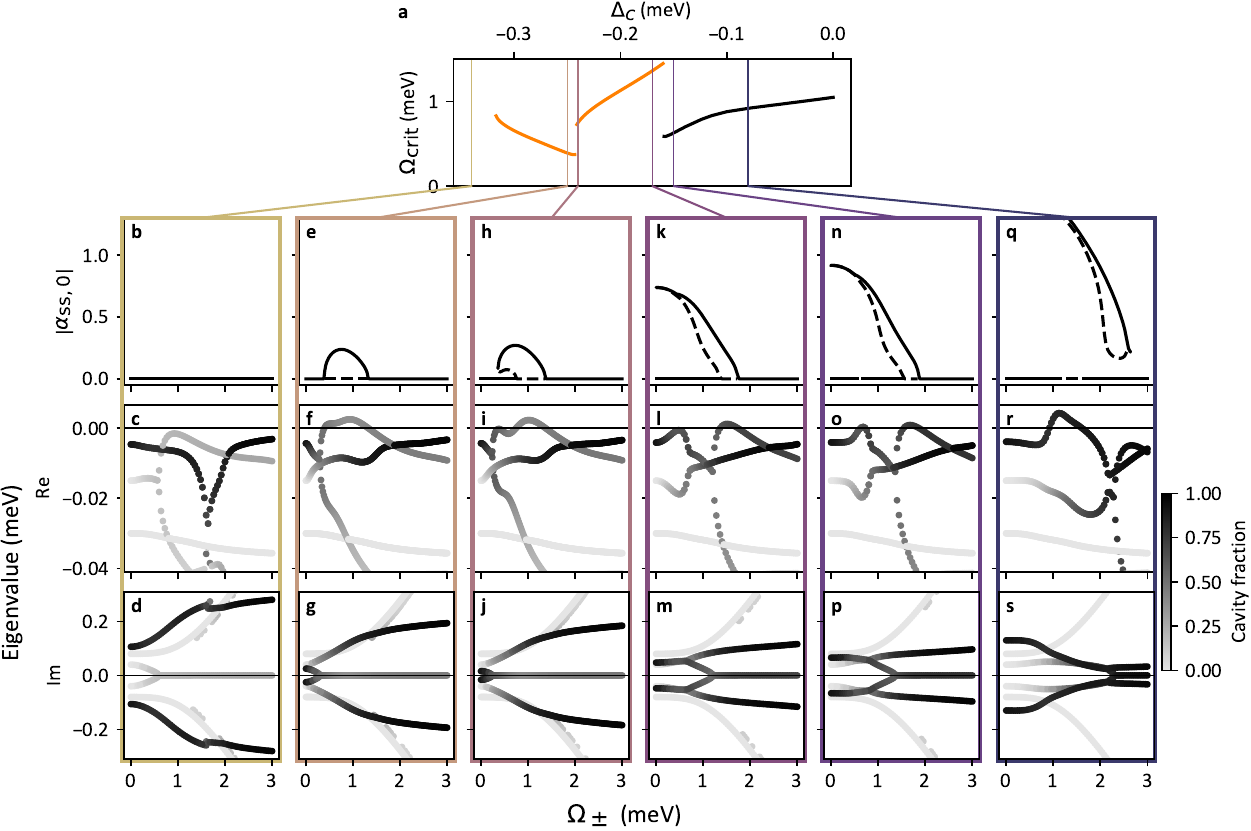}
    \caption{(a) Horizontal slice of the phase diagram of the full model, as shown in Fig.~\ref{fig:phase_diagram}(b), at $\Delta_{\text{ex}} = \qty{-34.5}{\meV}$, indicating the pump power at which the normal state goes unstable. The orange lines are due to a superradiant transition, while the black line is due to the lasing transition. (b-s) cavity amplitude bifurcation diagram and normal state eigenvalues at particular values of $\Delta_c$ from the slice in (a). Only the 6 nontrivial eigenvalues with the greatest real parts are shown. Parameter values are in Table~\ref{tab:parameters}, along with $\Omega_- = \Omega_+$ as well as (b-d) $\Delta_c = \qty{-0.34}{\meV}$, (e-g) $\Delta_c = \qty{-0.25}{\meV}$, (h-j) $\Delta_c = \qty{-0.24}{\meV}$, (k-m) $\Delta_c = \qty{-0.17}{\meV}$, (n-p) $\Delta_c = \qty{-0.15}{\meV}$, and (q-s) $\Delta_c = \qty{-0.08}{\meV}$.}
    \label{fig:horizontal_slice_eigs}
\end{figure}

In the main text, we showed the eigenvalues of the normal stationary state in the vicinity of the continuous phase transition to a superradiant state. In Fig.~\ref{fig:horizontal_slice_eigs} we show the eigenvalues of the normal stationary state as a function of pump strength at many values of $\Delta_c$, with all other parameters fixed. Figure~\ref{fig:horizontal_slice_eigs}(h-j) displays the eigenvalues for the bistable case shown in Fig.~\ref{fig:superradiant_bistability}, and Fig.~\ref{fig:horizontal_slice_eigs}(q-s) displays the eigenvalues for the lasing case shown in Fig.~\ref{fig:lasing_time_evolution}. In the superradiant cases, Fig.~\ref{fig:horizontal_slice_eigs}(e-m), there is only one real eigenvalue gaining a positive real part. The threshold for the superradiant transition rises as $\Delta_c$ increases above $\qty{-0.24}{\meV}$. At yet larger $\Delta_c$, a pair of complex-conjugate eigenvalues gain a positive real part at a pumping strength below the superradiant crossover, as shown in Fig.~\ref{fig:horizontal_slice_eigs}(n-p). This is the instability to lasing.

Note that the limit cycle shown in  Fig.~\ref{fig:lasing_time_evolution} at $\Omega_\pm = \qty{1}{\meV}$ does not correspond to any of the stationary state solutions found by the solver and shown in Fig.~\ref{fig:horizontal_slice_eigs}(q).  This is expected since the limit cycle is not a stationary state.  One may further note that in the limit cycle, the cavity field oscillates within a range $0.4 \lesssim |\alpha| \lesssim 0.6$.  These values fall between the stationary state solutions at $\left| \alpha_{\text{ss}, 0} \right|=0$ and $\left| \alpha_{\text{ss}, 0} \right| \simeq 1.25$. In the upper stationary state solution, the zero-frequency component of the exciton number expectation value nears half its Fock-space cutoff of $M=3$, indicating saturation and likely invalidating the solution. As a result, if the upper solution is spurious and the lower solution is unstable, then the system truly does not have any stable stationary states, indicating that its dynamics are not eventually periodic at period $T = 2\pi / \Delta\omega$.

\section{Finding eigenvalues corresponding to the dressed cavity and exciton}
\begin{figure}
    \centering
    \includegraphics[width=\textwidth]{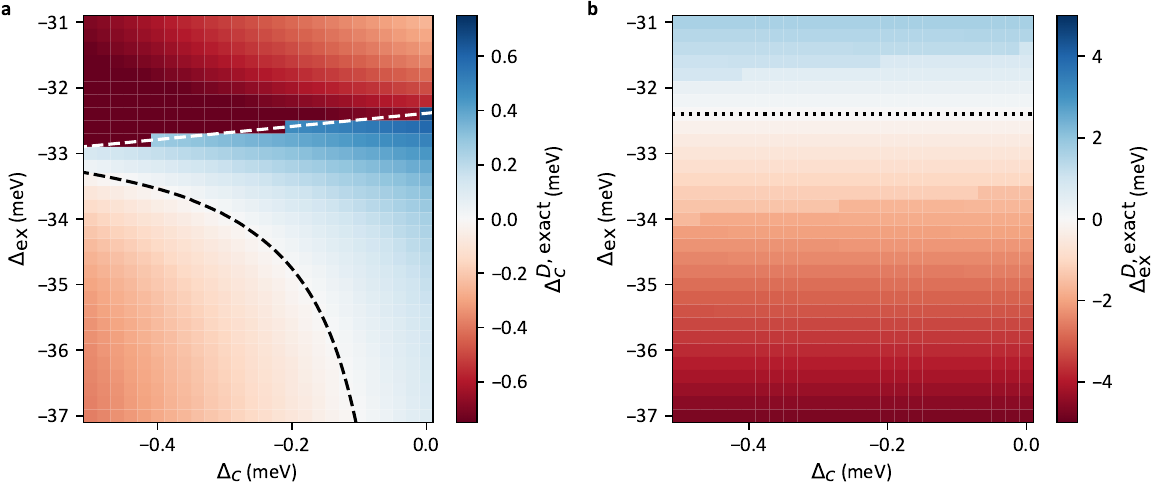}
    \caption{Imaginary parts of the eigenvalues corresponding to the (a) cavity and (b) exciton at zero pumping, yielding estimates of the dressed detunings. The lines are the same as in previous figures; dashed lines separate the regions where $\Delta_c^{D,\text{exact}}$ is positive (lower right) and negative (lower left and top), and the dotted line is where $\Delta_{\text{ex}}^{D,\text{exact}} = 0$. Parameters are in Table~\ref{tab:parameters}, along with $\Omega_\pm = 0$.}
    \label{fig:zero_pump_cavity_exciton_eigs}
\end{figure}

While the main text gives explicit perturbative expressions for the dressed exciton and cavity detunings, these quantities may be found more accurately by finding the eigenvalues and eigenvectors of the linearised system at zero pumping, $\Omega_\pm = 0$, and identifying the relevant eigenvalues. 
For the exciton this means taking the eigenvalue corresponding to the eigenvector with the greatest weight for coherence between the vacuum to the one-exciton state. Note that the linearised system consists of the cavity photon field and the vectorised density matrix $\rho$, so the excitonic mode corresponds to a coherence. A similar approach finds the photon mode. The imaginary parts of the resulting eigenvalues provide precise numerical estimates of the dressed exciton and cavity detunings, while the real parts are estimates of the dressed linewidths. We refer to these numerically-found dressed detunings as $\Delta_{\text{ex}}^{D,\text{exact}}$ and $\Delta_c^{D,\text{exact}}$ and present them in Fig.~\ref{fig:zero_pump_cavity_exciton_eigs}. This procedure determines whether the exciton and/or cavity are effectively red- or blue-detuned, and it is much faster than handling nonzero pumping power for two reasons. Firstly, the vacuum is a stationary state, so one does not need to do the root-finding for $\alpha_{\text{ss}, n}$. And secondly, there is no longer explicit time dependence, so rather than computing the time-ordered exponential of $\hat{A}$ as in Eq.~\eqref{eq:monodromy}, one may diagonalise $\hat{A}$ directly.

\section{Further details on the threshold power estimate}
This section provides further details on several topics affecting the estimate of critical laser power in the main text.

\subsection{Dressed cavity linewidth and effects of exciton matrix element}
In the main text we obtained an expression for the threshold pump power assuming that the cavity mode is so strongly dressed by mixing with excitons that the cavity linewidth is proportional to $\Gamma_{\text{ex}}$, resulting in the threshold expression of Eq.~\eqref{eq:threshold-power}. 
However, as also discussed in the main text, there are three different regimes of cavity linewidth that may arise:
\begin{equation}
\kappa^D \simeq
    \begin{cases}
        \kappa & 
        g_0^2 {N} 
        \ll 
        \kappa \Gamma_{\text{ex}} \\
        \displaystyle
        \frac{g_0^2 N \Gamma_{\text{ex}} }{2 (\Deltaexdressed-\Delta_c)^2} &
        \kappa \Gamma_{\text{ex}}
        \ll 
        g_0^2 {N} 
        \ll 
        (\Deltaexdressed-\Delta_c)^2
        \\
        {\Gamma_{\text{ex}}}/{4}
        &
        (\Deltaexdressed-\Delta_c)^2
        \ll 
        g_0^2 {N}.
    \end{cases}
\end{equation}
The results in the main text assume the third of these regimes.
The most realistic experimental parameters place the system on the border between the second and third regimes.
In the following we will discuss the form of the critical laser power in each of these regimes, and comment in particular on how the threshold depends on exciton matrix element $e_{\text{ex}}$ and cavity field strength $E_c$ in each regime.

To consider how the threshold power varies in different regimes, we focus on how the threshold power depends on light-matter coupling and optical linewidths.  (Other factors, such as those depending on phonon coupling, remain the same in all three regimes).  The threshold power can then be considered to take the form $P_{\text{crit}} \propto \Gamma_{\text{ex}}^2 \kappa^D / \left( e_{\text{ex}}^4 E_c^2 \right)$.

The result in Eq.~\eqref{eq:threshold-power} can be seen to follow from the expression above by taking $\kappa^D={\Gamma_{\text{ex}}}/{4}$, as occurs in the extremely high-coupling regime.  
In this regime, increasing the cavity electric field reduces the threshold power as might be expected.  The dependence on exciton matrix element is however more subtle.  This is because the radiative part of the exciton linewidth $\Gamma_{\text{ex}}$ will depend on the same matrix element.  Thus, if the exciton linewidth is purely radiative then $\Gamma_{\text{ex}} \propto e_{\text{ex}}^2$ and greater light-matter coupling $e_{\text{ex}}$ \emph{increases} the threshold power.

In the extremely low-coupling regime $g_0^2 {N} \ll \kappa \Gamma_{\text{ex}}$, the dressing is minimal and $\kappa^D \simeq \kappa$.   In this regime increasing the cavity electric field again reduces the threshold power, however increasing $E_c$ will also drive one out of this low-coupling regime.  In terms of dependence on exciton matrix element, one may note that in this case, if we again assume the exciton linewidth is purely radiative, then the factors of $e_{\text{ex}}$ entirely cancel, so the scaling is $P_{\text{crit}} \propto  \kappa / E_c^2$.

In the intermediate coupling regime, the dressed cavity linewidth is $\kappa^D \propto g_0^2 N \Gamma_{\text{ex}} / \left(\Deltaexdressed\right)^2$ (where we have assumed $\Deltaexdressed \gg \Delta_c$ as discussed in the main text).
If we further take $\Deltaexdressed \propto \Gamma_{\text{ex}}$ (also as discussed in the main text), the dressed cavity linewidth is inversely proportional to the exciton linewidth as $\kappa^D \propto g_0^2 N / \Gamma_{\text{ex}}$. Physically this is because larger exciton linewidth forces one to detune further away, reducing the light-matter coupling, and therefore reducing the dressing. In this regime, the threshold pump power scales as $P_{\text{crit}} \propto \Gamma_{\text{ex}} / e_{\text{ex}}^2$. 
Here the cavity field strength becomes irrelevant.  If we further take the exciton linewidth to be purely radiative, all factors relating to exciton matrix element also cancel. In this case, the only remaining material property that affects threshold power is the phonon linewidth.

\subsection{Emitter density}

The model studied in the main text is based on localised emitters, where each ``site'' can host zero or one excitons.  To estimate physical critical laser powers, we want to connect such a model to the properties of 2D materials, in which excitons are instead delocalised bosonic excitations. 
This section provides further details of how to establish this connection, and thus further motivates the parameter estimates given in the main text.

We start by writing the creation operator for a delocalised  exciton in a 2D material.  For an exciton with centre-of-mass momentum $\mathbf{K}$ and relative electron-hole state $\nu$ we can write:
\begin{equation}
    \hat{X}^\dagger_{\nu, \mathbf{K}} = \sum_{\mathbf{k}} \psi^{}_{\nu, \mathbf{k}} \hat{c}^\dagger_{c, \mathbf{k} + \gamma_c \mathbf{K}} \hat{c}^{}_{v, \mathbf{k} - \gamma_v \mathbf{K}},
\end{equation}
where $\hat{c}_{\alpha=(c, v),\mathbf{k}}$ are the annihilation operators for carriers in the conduction and valence band, respectively; $\gamma_\alpha = \left| m_\alpha \right| / \left( |m_c| + |m_v| \right)$ is the effective mass fraction of species $\alpha$; and $\psi_{\nu, \mathbf{k}}$ is the momentum-space wavefunction of exciton state $\nu$.  This wavefunction is normalied by
$ \sum_{\mathbf{k}} \left| \psi_{\nu, \mathbf{k}} \right|^2 = 1$, and in the following we will make use also of its real-space representation $\psi_\nu(\mathbf{r}) = \sum_{\mathbf{k}} \psi_{\nu,\mathbf{k}} e^{i \mathbf{k}\cdot\mathbf{r}}/\sqrt{A}$ where $A$ is the quantisation area.
Exciton states created on top of the material vacuum $\ket{0}$ of a filled valence band and empty conduction band can be written as $\ket{\nu, \mathbf{K}} = \hat{X}^\dagger_{\nu, \mathbf{K}} \ket{0}$.  
While we will ultimately only consider the lowest-lying state $\nu = \text{1s}$ at $\mathbf{K} = \mathbf{0}$ in what follows, it is illustrative to temporarily retain the indices.

To consider exciton-photon coupling, we
start from the first-quantised Hamiltonian of the electric dipole interaction~\cite{Haug2004qto}, $\hat{H}_I = -e \mathbf{\hat{r}} \cdot \mathbf{\hat{E}}$, for an applied electric field $E \pmb{\hat{\varepsilon}}$ polarised in the $\pmb{\hat{\varepsilon}}$ direction. Note that the hat in $\mathbf{\hat{r}}$ refers to its operator nature and not that it is a unit vector, while $\pmb{\hat{\varepsilon}}$ is a unit vector.  This light-matter interaction term can then be written in terms of second-quantised electron creation and annihilation operators as:
\begin{equation}
    \hat{H}_I = -\sum_{\mathbf{q}, \mathbf{k}} \left( d_{cv, \mathbf{k + q}, \mathbf{k}} E_{\mathbf{k}} \hat{c}^\dagger_{c, \mathbf{k} + \mathbf{q}} \hat{c}_{v, \mathbf{q}} + \text{H.c.} \right),
\end{equation}
where 
$d_{cv,\mathbf{q^\prime},\mathbf{q}} = e \matrixel{0}{ \hat{c}^\dagger_{c, \mathbf{q^\prime}} \mathbf{\hat{r}} \cdot \pmb{\hat{\varepsilon}} \hat{c}_{v,\mathbf{q}} }{0}$ is the the interband dipole matrix element, which is an intensive material property.
Note that in the dipole approximation, intraband terms can be assumed to vanish.  Using the completeness of the exciton states, this interaction can then be further rewritten as:
\begin{equation}
\hat{H}_I = -\sum_{\nu, \mathbf{K}} \left( M_{\nu, \mathbf{K}} E_{\mathbf{K}} \hat{X}^\dagger_{\nu, \mathbf{K}} + \text{H.c.}\right), \qquad
    M_{\nu, \mathbf{K}} = e \matrixel{\nu, \mathbf{K}}{\mathbf{\hat{r}} \cdot \pmb{\hat{\varepsilon}}}{0} = \sum_{\mathbf{q}} d_{cv, \mathbf{q}, \mathbf{q}} \psi^*_{\nu, \mathbf{q}}.
\end{equation}
In writing this we have used the approximation that the typical centre-of-mass momentum, $\mathbf{K}$, is small, so that we can neglect the dependence of $d_{cv, \mathbf{q}+\mathbf{K}, \mathbf{q}}$ on $\mathbf{K}$.
This expression contains the subtlety that  $M_{\nu, \mathbf{K}}$ is not an intensive quantity: The number of terms in the sum over $\mathbf{q}$ scales with the area $A$, while the normalisation of $\psi_{\nu,\mathbf{q}}$ implies it scales as $1/\sqrt{A}$.

We now focus on the 1s exciton at $\mathbf{K} = \mathbf{0}$. For illustration, if one assumes that $\psi^*_{\text{1s}, \mathbf{q}}$ is sharply peaked around $\mathbf{q} = \mathbf{0}$ and $d_{cv}$ varies little throughout this region,  one then obtains $M_{\text{1s}, \mathbf{0}} = d_{cv, \mathbf{0}, \mathbf{0}} \psi^*_{\text{1s}}(\mathbf{r} = \mathbf{0}) \sqrt{A}$, scaling with $A$ as anticipated above.  Based on this we introduce the intensive exciton matrix element by defining $e_{\text{ex}} \equiv M_{\text{1s}, \mathbf{0}} / \sqrt{A}$. The interaction Hamiltonian between $1s$ excitons and light now reads, in position space:
\begin{equation}
    \label{eq:HI_Xr}
    \hat{H}_I = -e_{\text{ex}} \int d^2\mathbf{r}\, \left[ E(\mathbf{r}) \hat{X}^\dagger(\mathbf{r}) + \text{H.c.} \right],
\end{equation}
where $\hat{X}(\mathbf{r}) = \sum_{\mathbf{K}} \hat{X}_{\mathbf{K}} e^{i \mathbf{K}\cdot\mathbf{R}}/\sqrt{A}$, so that
$\hat{X}^\dagger(\mathbf{r}) \hat{X}(\mathbf{r})$ has units of inverse area.  Note that this expression assumes $M_{\text{1s}, \mathbf{K}}$ depends weakly on $\mathbf{K}$, so that the $\mathbf{K}$ dependence of the matrix element can be neglected.

To go from Eq.~\eqref{eq:HI_Xr} to the localised emitter model, we then discretise space into $N$ cells, so that $\int d^2\mathbf{r} \Rightarrow n^{-1} \sum_{j = 1} ^N$ and $\hat{X}(\mathbf{r}_j) \Rightarrow \sqrt{n} \hat{X}_j$ for emitter density $n = N / A$. Thus excitons $\hat{X}_j$ in the independent-oscillator model couple to electric fields $E$ via $\left( e_{\text{ex}} / \sqrt{n} \right) E$, and one must choose a suitable $n$.
As discussed in the main text, we use the inverse unit cell area for $n$.

\end{document}